\documentclass[a4paper,fleqn,usenatbib]{mnras}

\usepackage[T1]{fontenc}
\usepackage{ae,aecompl}

\usepackage{subfigure}
\usepackage{graphicx}	
\usepackage{amsmath}	
\usepackage{amssymb}	
\usepackage{txfonts}
\usepackage{ulem}
\usepackage{ulem}

\title[Sampling the IMF in Cosmological Simulations]{Roll of the Dice: A Stochastically Sampled IMF Alters the Stellar Content of Simulated Dwarf Galaxies}

\author[E. Applebaum et al.]{Elaad Applebaum,$^{1}$\thanks{E-mail: applebaum@physics.rutgers.edu}
Alyson M. Brooks,$^{1}$
Thomas R. Quinn,$^{2}$
\newauthor and Charlotte R. Christensen$^{3}$
\\
$^{1}$Department of Physics and Astronomy, 
Rutgers, The State University of New Jersey, 
136 Frelinghuysen Road, 
Piscataway, NJ 08854, USA\\
$^{2}$Department of Astronomy, 
University of Washington, 
Box 351580, 
Seattle, WA 98195, USA\\
$^{3}$Physics Department, 
Grinnell College, 
1116 Eighth Avenue, 
Crinnell, IA 50112, USA
}

\date{Accepted XXX. Received YYY; in original form ZZZ}

\pubyear{2018}

\begin{document}
\label{firstpage}
\pagerange{\pageref{firstpage}--\pageref{lastpage}}
\maketitle

\begin{abstract}
Cosmological simulations are reaching the resolution necessary to study ultra-faint dwarf galaxies. Observations indicate that in small populations, the stellar initial mass function (IMF) is not fully populated; rather, stars are sampled in a way that can be approximated as coming from an underlying probability density function. To ensure the accuracy of cosmological simulations in the ultra-faint regime, we present an improved treatment of the IMF. We implement a self-consistent, stochastically populated IMF in cosmological hydrodynamic simulations. We test our method using high-resolution simulations of a Milky Way halo, run to $z=6$, yielding a sample of nearly 100 galaxies. We also use an isolated dwarf galaxy to investigate the resulting systematic differences in galaxy properties. We find that a stochastic IMF in simulations makes feedback burstier, strengthening feedback, and quenching star formation earlier in small dwarf galaxies. For galaxies in halos with \mbox{mass $\lesssim10^{8.5}$~M$_\ssun$}, a stochastic IMF typically leads to lower stellar mass compared to a continuous IMF, sometimes by more than an order of magnitude. We show that existing methods of ensuring discrete supernovae incorrectly determine the mass of the star particle and its associated feedback. This leads to overcooling of surrounding gas, with at least ${\sim}$10 per cent higher star formation and ${\sim}$30 per cent higher cold gas content. Going forward, to accurately model dwarf galaxies and compare to observations, it will be necessary to incorporate a stochastically populated IMF that samples the full spectrum of stellar masses.
\end{abstract}




\section{Introduction} \label{sec:intro}

Prior to the last decade, dwarf galaxies posed a long-standing challenge to galaxy formation models within the context of the cold dark matter (CDM) paradigm. In recent years, however, enormous progress has been made in simulating these low-mass systems, owing to increasing resolution and careful modeling of the baryonic processes involved. For example, repeated fluctuations of the gravitational potential well of a halo due to supernova-induced gas outflows have been shown to flatten the central dark matter profile \citep[e.g.][]{Read2005, Mashchenko2008, Governato2010, Pontzen2012, DiCintio2014, Read2016, Fitts2017}, alleviating tension with observations \citep[e.g.][]{Simon2005, DeBlok2008, KuzioDeNaray2008}. Supernova-driven outflows also remove low-angular momentum gas, leading to the creation of bulgeless dwarf galaxies \citep{Governato2010, Brook2011, Teyssier2013}. Similar effects together with enhanced tidal stripping within the parent halo successfully reduce the central densities of the most massive Milky Way satellites \citep[e.g.][]{Zolotov2012, Brooks2014, Wetzel2016, Tomozeiu2016, Sawala2016, Garrison-Kimmel2019}, solving the well-known ``too big to fail'' problem \citep{Boylan-Kolchin2011, Boylan-Kolchin2012}. Destruction of satellite halos from interactions with the central disk potential reduces the theoretically expected number of MW satellites \citep[e.g.][]{Brooks2013, Wetzel2016, Sawala2016, Garrison-Kimmel2019}, solving the ``missing satellites'' problem \citep{Klypin1999, Moore1999}. Proper consideration of the biases introduced when comparing observations to theoretical results also brings theory into agreement with observations \citep{Brooks2017}.

With the recent successes in modeling dwarf galaxies, various groups are pushing the limits of resolution even further, into the ultra-faint dwarf (UFD) galaxy range (M$_\mathrm{star}\lesssim 10^5$~M$_\ssun$; e.g. \citealt{Munshi2013, Wheeler2015, Munshi2017, Munshi2019}; Garrison-Kimmel~in~prep).

Despite significant differences in feedback models and simulation details, many groups have succeeded in reproducing realistic dwarf galaxies. This is largely a result of galaxy self-regulation; i.e. global galaxy properties are robust to the details of star formation and feedback, since increased feedback suppresses future star formation \citep{Saitoh2008, Shetty2008, Hopkins2011, Christensen2014b, Agertz2015, Benincasa2016, FIRE-2}. As simulations approach the UFD regime, however, the self-regulation of galaxies breaks down. \citet{Munshi2019} demonstrated this conclusively, finding that different star formation prescriptions predict different numbers of UFD galaxies. Further, the interplay of supernova feedback and reionization with the gas in these low-mass halos leads to diverging star formation behavior at early times. Thus, without a concerted effort to study subgrid models in the low-mass galaxy regime, the predictive power of cosmological simulations will diminish.

In this work we investigate another physical prescription that can alter UFD simulation results: the realistic sampling of stars and its impact on subsequent feedback. Since the high-mass stars providing the bulk of stellar feedback are relatively uncommon, nuances in choice of sampling may not ``average out'' in UFDs. It is therefore important to correctly determine the stellar masses within a population; this is done through the use of the stellar initial mass function (IMF), which describes the number distribution of stars as a function of their birth mass.

Estimates of the IMF in large stellar populations have shown remarkable consistency. Parameterizations generally find a steep power-law slope for more massive stars consistent with the original \citet{Salpeter1955} estimate, with a ``knee'' at m$_\mathrm{star}\sim$~M$_\ssun$, below which the distribution involves a shallower decline \citep[e.g.][]{Kroupa1993, Kroupa2001, Chabrier2003}. While theoretical expectations predict systematic variation in the IMF with environment \citep{Kroupa2013}, in large resolved stellar populations there is limited evidence of deviations from the universal IMF \citep{Bastian2010}. While several recent observations suggest systematic variations \citep[e.g.][]{vanDokkum2010,Cappellari2012,Conroy2012,Kalirai2013,Geha2013,Gennaro2018}, there is neither consensus on their significance nor a clear physical driver for their variation, with dominant candidates including metallicity \citep[e.g.][]{MartinNavarro2015, Gennaro2018} and velocity dispersion \citep[e.g.][]{LaBarbera2013, Spiniello2014, Rosani2018}

Despite the general success of the IMF formalism in describing galaxies and large stellar populations, in small populations it is clear that the current IMF formalism is insufficient. The inherently discrete nature of stars makes a continuous description unrealistic. To find a better description, a variety of observations can be used. For example, there exists an average relationship between the mass of an embedded star cluster and the mass of the most massive star residing in the cluster, such that more massive stars tend to reside in more massive clusters (\citealt{Weidner2010,Weidner2013a}; see also \citealt{Cervino2013} for a detailed discussion). Other observations have shown lower values of H$\alpha$- or H$\beta$-to-FUV luminosity ratios in galaxies with low star formation rates (SFRs), indicating a relative dearth of very high-mass stars \citep{Meurer2009, Lee2009, Lee2016}.

It follows from the above observations that there is a tendency for fewer high-mass stars to form in small, low-SFR populations. Observations of SFR indicators can be explained via bursty star formation histories \citep{Weisz2012, Guo2016, Emami2019}, but IMF sampling effects may also contribute \citep[e.g.][]{Pflamm-Altenburg2009, Fumagalli2011, Eldridge2012}. 

There are two broad theories of how the IMF should be sampled to explain these observations. The first is the integrated galactic IMF (IGIMF), presented in \citet{Kroupa2003,Weidner2006,Weidner2010,Weidner2013b,Yan2017}, which assumes a deterministic relationship between the mass of a star cluster and the stellar mass distribution within it. The other predominant explanation for the observations is that the IMF is sampled randomly \citep[e.g.][]{Elmegreen2006, Corbelli2009, Calzetti2010, Fumagalli2011, Andrews2013, Andrews2014}. In every star formation event, stars form in a way approximated as being drawn from an underlying probability density function---the universal IMF. Since the probability of forming high-mass stars is rare, in small populations there is an \textit{average} tendency for massive stars to form in massive clusters, mimicking the proposed IGIMF.

The only restriction in stochastically sampling the IMF is that a star cannot form with a greater mass than the cluster in which it resides (i.e. stars cannot form with more mass than their available gas reservoirs). Until recently, star particles in cosmological simulations were large enough to ignore all the nuances of IMF sampling, and stellar feedback was calculated by treating star particles as a simple stellar population with a uniform, continuous IMF. With sufficiently massive star particles, this was a reasonably accurate approximation. At lower particle masses, however, the above model has proven increasingly unrealistic. Further, star particles are small enough such that not only is a uniform IMF no longer consistent with observations, but a naive calculation of supernova counts per time step yields fractions of supernovae exploding \citep{Revaz2016}.

In small galaxies, the credibility of simulated results depends upon proper treatment of the IMF. \citet{Carigi2008} demonstrated that a stochastically sampled IMF does not converge to the underlying continuous IMF until M$_\mathrm{star}\sim10^5$~M$_\ssun$ (see also \citealt{Hernandez2012}). Cosmological galaxy simulations are now pushing to high enough resolution to study stellar populations in the ultra-faint regime; at these scales, not only individual star particles but also entire galaxies will have incompletely sampled IMFs.

Limited work has been done in investigating the impacts of IMF sampling within cosmological simulations. In post-processing, \citet{Sparre2017} used the SLUG code \citep{SLUG1, SLUG2, SLUG3} to show how a stochastic IMF increases the scatter of dwarf galaxies' H$\alpha$-to-FUV ratios, but the simulation itself assumed a fully populated IMF. \citet{Revaz2016} studied the effects on IMF sampling on stellar chemical abundances in isolated dwarf galaxies, and found that a continuous IMF becomes unrealistic at star particle masses below ${\sim}10^5$~M$_\ssun$. They further found that regardless of sampling method, the combined IMF of multiple star particles together will be undersampled below particle masses of ${\sim}10^3$~M$_\ssun$. Bracketing the case of a stochastic IMF, \citet{Hensler2017} compared a truncated and a filled IMF in simulations of dwarf galaxies, and found that truncation suppresses the self-regulation of star formation. Several cosmological simulations have discussed or incorporated methods of discretizing stellar feedback from Type II supernovae \citep[e.g.][]{Stinson2010, Agertz2013, Hopkins2014, FIRE-2, Rosdahl2018}. The most common method has been to decide whether or not a star explodes by drawing from a binomial or Poisson distribution derived from an average measure of supernova rates. The drawbacks of this method are discussed in Section~\ref{sec:quant}. \citet{su2018} took first steps in investigating IMF sampling and stochastic effects more closely; they found a dramatic decrease in star formation when discretizing their supernovae compared to continuous energy injection. However, their model does not sample the full range of masses in the IMF and still calculates feedback by drawing from a Poisson distribution.

More work has been done in high resolution simulations that do not model cosmological contexts. \citet{Grudic2019} used the same methodology as \citet{su2018} on molecular cloud scales and found similar results. \citet{Sormani2017} introduced a method based on discretizing stars into mass bins which are then Poisson sampled. Other groups \citep[e.g.][]{Gatto2017, Geen2018} separate their IMF into high- and low-mass regimes, and stochastically sample only within the high-mass regime. We note that some recent simulations of very small, isolated dwarf galaxies now track the evolution of individually sampled stars \citep{Hu2017, Emerick2019} and recent work with isolated Milky Way-mass galaxies includes a stochastically populated IMF within star particles \citep{Fujimoto2018}, but this has never previously been attempted in cosmological simulations.

In this paper, we present a new prescription for star formation that stochastically samples the full spectrum of masses in the IMF and individually tracks the evolution of high-mass stars within them. This methodology ensures conservation of mass and self-consistency of radiative and supernova feedback. We discuss the simulations and the sampling method in Section~\ref{sec:methods}, and compare to existing discretization methods. In Section~\ref{sec:results} we demonstrate the effects of improved IMF sampling. In Section~\ref{sec:discussion} we discuss implications of this sampling method for future observational predictions. We conclude in Section~\ref{sec:summary}.

\section{Methods}\label{sec:methods}

We implement a new stochastic IMF treatment for star particles in our simulations. The updated sampling changes the stellar mass distribution from a smooth IMF to a set of discrete stars; these stars are then used to calculate supernova explosions, metal production, and high-energy radiation output. We emphasize that the actual feedback mechanisms remain unaltered; what changes in the new recipe is the timing and quantity of feedback each star particle produces.

\subsection{Simulations}\label{sec:simulations}

To test the new IMF prescription, we ran cosmological zoom-in simulations of a Milky Way-mass galaxy with and without the stochastic IMF. These were run until immediately after reionization, by which time ultra-faint dwarf galaxies---those with M$_\mathrm{star}\lesssim10^5$~M$_\ssun$, and therefore unconverged IMFs---have formed most or all of their stars \citep{Brown2014, Weisz2014b}. Since the same dark matter particles form the same halos in simulations with and without a stochastic IMF, we were able to match galaxies that formed in corresponding halos between runs. To ensure our results are independent of any intrinsic scatter in the cosmological runs (and allow a finer time resolution of outputs), we also used an isolated dwarf galaxy. For both treatments of the IMF, we ran the same $10^9$~M$_\ssun$ halo 50 times, and compared the ensemble behavior.

\begin{table*}
    \centering
    \caption{Properties of the cosmological simulations used in this work, including particle masses, the expectation value of core collapse supernovae per star particle, and the softening length.}
    \begin{tabular}{cccccc}\hline
        Simulation & M$_\mathrm{dark}$ & M$_\mathrm{gas}$ & M$_\mathrm{star}$ & $\langle\mathrm{M}_\mathrm{SNII}\rangle$ & Softening \\
         & [M$_\ssun$] & [M$_\ssun$] & [M$_\ssun$] & & Length [pc] \\\hline
         Elena (Milky Way) & $1.8\times10^4$ & $3.3\times10^3$ & 990 & 9.9 & 87 \\
         Isolated $10^9$~M$_\ssun$ & $1.0\times10^4$ & $1.4\times10^3$ & 420 & 4.2 & 21
         \\
         \hline
    \end{tabular}
    \label{tab:properties}
\end{table*}

The relevant properties of each simulation can be found in Table~\ref{tab:properties}. The cosmological simulations used in this work were selected from a uniform resolution, dark matter-only simulation of 50~Mpc per side, run with Planck cosmological parameters \citep{Planck2016}. A region around the selected halo in these simulations was rerun at higher resolution using the ``zoom-in'' technique \citep{Katz1993}. The halo used in this work was selected to resemble the Milky Way at $z=0$, and
is one of the halos of the DC Justice League suite of simulations \citep{Bellovary2019}, nicknamed ``Elena.'' We use a gravitational force softening length of 87~pc and equivalent resolution to a $6144^3$ grid. The dark matter particle masses are $1.8\times10^4$~M$_\ssun$, the gas particles begin with $3.3\times10^3$~M$_\ssun$, and the star particles form with a mass of 994~M$_\ssun$. Versions of the DC Justice League suite at this resolution are being run to the present day, and will constitute the highest resolution Milky Way simulations to date.

The initial conditions for the $10^9$~M$_\ssun$ isolated dwarf galaxy have been described previously \citep{Kaufmann2007, Stinson2007, Christensen2010}. In short, the initial conditions consist of an equilibrium halo with a Navarro-Frenk-White concentration of $c=8$. Dark matter velocities were determined via the Eddington inversion method of \citet{Kazantzidis2004}. Gas particles were assigned temperatures to ensure hydrostatic equilibrium before cooling, and were given a uniform rotational velocity corresponding to a spin parameter ${\sim}0.04$. Dark matter particles within the virial radius have a mass of $1.0\times10^4$~M$_\ssun$, gas particles have a mass of $1.4\times10^3$~M$_\ssun$, and star particles from with 425~M$_\ssun$. The force softening length is 0.1\% the virial radius, or 21~pc.

The stochastically populated IMF is incorporated into the \mbox{N-body} + Smoothed Particle Hydrodynamics (SPH) code \textsc{ChaNGa} \citep{Menon2015}, a fully cosmological simulation code which includes physics from the \textsc{Gasoline2} code \citep{Wadsley2017}, but utilizes the \textsc{charm++} runtime system for dynamic load balancing to efficiently scale up to thousands of cores. All simulations discussed in this work smooth over 32 nearest neighbor gas particles.

As discussed above, feedback from high-mass stars is crucial for modeling realistic galaxies. In this work we use the ``blastwave'' supernova feedback mechanism described in detail in \citep{Stinson2006}, whereby mass, energy, and chemically enriched material are deposited into neighboring gas when a massive star dies as a Type II supernova. With the existing continuous IMF, the minimum and maximum stars that explode in a given time step are calculated based on the stellar lifetime parameterizations of \citet{Raiteri1996}. The number and mass in supernovae are then determined by integrating along the IMF between these stellar masses. We note that the default time step for calculating feedback and star formation in \textsc{ChaNGa} is 1~Myr. We deposit $1.5\times10^{51}$ erg per supernova\footnote{Rates of supernova energy deposition were determined using the parameter optimization technique described in \citet{Tremmel2017} and \citet{Anderson2017}.} among gas particles within the blast radius calculated using \citet{McKee1977}, then gas cooling is shut off until the end of the blastwave's snowplow phase. We assume stars between 8 and 40 M$_\ssun$ explode as supernovae, while more massive stars collapse into black holes. Future work will incorporate a stochastic IMF using the ``superbubble'' feedback mechanism \citep{Keller2014}. Results from \textsc{ChaNGa} and \textsc{Gasoline} with blastwave feedback have been used to reproduce a variety of observations, including the stellar mass-halo mass relation \citep{Munshi2013, Munshi2017}, the mass-metallicity relation \citep{Brooks2007}, the baryonic Tully-Fisher relation \citep{Christensen2016, Brooks2017}, the abundance of Damped Lyman $\alpha$ systems \citep{Pontzen2008}, and the properties of dwarf Spheroidal Milky Way satellites \citep{Brooks2014}. These models also produced the first simulated cored dark matter density profiles and bulgeless disk galaxies \citep{Governato2010, Brook2011, Governato2012}.

In addition to feedback from core collapse supernovae, \textsc{ChaNGa} incorporates metal cooling and diffusion in the ISM \citep{Shen2010}, a time-dependent UV background \citep{Haardt2012}, Type Ia supernovae, mass loss in stellar winds, and metal enrichment \citep{Stinson2006}, and supermassive black hole formation, growth, and feedback \citep{Tremmel2015, Tremmel2017}.

\textsc{ChaNGa} includes a star formation recipe based on the local abundance of molecular hydrogen \citep[H$_2$;][]{Christensen2012}. This scheme includes calculations for the formation of H$_2$, shielding from dissociative Lyman Werner (LW) radiation, and production of LW photons from high-mass stars. LW photon production from star particles is calculated using \textsc{Starburst99} \citep{Leitherer1999}; for star particles represented by a uniform IMF, as done until now, the calculation is based on a single-age, simple stellar population.

Halos in the cosmological simulations are identified using \textsc{Amiga's Halo Finder} \citep{Gill2004, Knebe2009}. Halos are defined as the radius within which the density reaches a redshift-dependent overdensity criterion using the approximation of \citet{Bryan1998}\footnote{At $z=0$, the overdensity compared to the critical density is $\rho/\rho_c\approx100$.}. Virial radius and halo mass are defined according to this overdensity.

\subsection{Stochastically populated IMF}\label{sec:stoch}

Ideally, no approximations would be necessary in cosmological simulations and every individual star would be tracked within a given stellar population. However, tracking hundreds of individual stars within each of millions of star particles would be computationally prohibitive. Any approximation, then, should be guided by two considerations: 1) we wish to preserve the highest accuracy for the individual stars whose feedback has the greatest impact on galaxy evolution, and 2) we wish to preserve the highest accuracy for individual stars that are rarest, and therefore are most altered by the approximations of a continuous, universal IMF. Therefore, we strive to maintain highest accuracy in the high-mass component; specifically, stars that eventually release energy as Type II supernovae.

Due to the nature of random sampling, the mass of a simulated stellar population cannot be predetermined if using a stochastically populated IMF; rather, only an estimate can be known a priori \citep[see, e.g.][]{Cervino2013}. Since in our simulations we form star particles of a given mass, we must use an algorithm to stochastically populate stars in a way that reaches the desired mass of our population.

When stochastically sampling from the IMF, the IMF is treated as a probability density function, so that its area is normalized to one but its form is otherwise unchanged. The algorithm we adopt is the stop-nearest method \citep[e.g.][]{SLUG1, Eldridge2012}. With this method, stars are drawn from the IMF until the desired mass is first exceeded. Then, the last star drawn is either kept or discarded, depending on whether the total mass is closer with or without its inclusion.

The methodology, then, for our stochastically populated IMF is as follows:
\begin{enumerate}
    \item If we determine a star particle forms, the formation mass is the target mass,
    \item Following the stop-nearest method, we draw stars from the IMF until we pass our target mass threshold, then either keep or discard the last star based on which brings us closer to the target mass,
    \item We discard all stars below a cutoff mass\footnote{Our fiducial cutoff mass is 8 M$_\ssun$. Appendix~\ref{app:cutoff} discusses what would happen if we raised the value of the cutoff.} and reapproximate the low-mass stars as a continuous IMF, normalized such that the total mass of the star particle is the target mass,
    \item For feedback dependent on stellar masses above the cutoff, we use the individual high mass stars to calculate the timing and quantity. For a cutoff of 8~M$_\ssun$, all energy and metals from Type II supernovae and LW photon production are calculated discretely; Type Ia supernovae and stellar winds are calculated as in the case of a continuous IMF, since they come from the stellar mass range approximated as a continuous distribution.
\end{enumerate}

This methodology shares similarities with that of, e.g., \citet{Gatto2017}; however, we sample stars over the entire mass range of the IMF rather than just among the high-mass end, to avoid imposing a restriction on either the mass in high-mass stars or the number. We thus allow for substantially more variation in the mass and number of high-mass stars per star particle. The numerical implications of this method are discussed briefly in Appendix~\ref{app:numerical}.

\begin{figure*}
    \centering
    \includegraphics[width=0.85\textwidth]{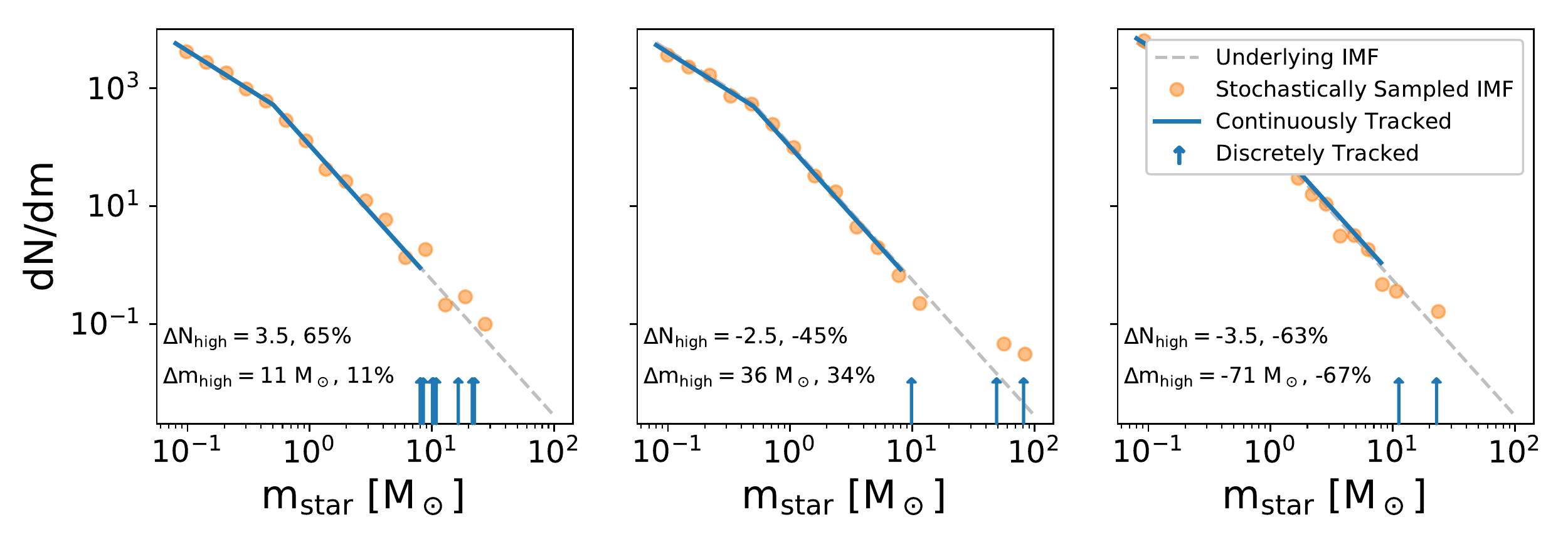}
    \caption{Three realizations of a stochastically populated 500 M$_\ssun$ star particle, following the methodology of Section~\ref{sec:stoch}. The grey dashed line shows the universal underlying IMF of \citet{Kroupa2001}, scaled to the population mass. The orange circles represent the IMF recovered from stochastically drawing from the entire IMF, to which we then apply a continuous/discrete cutoff. The solid blue line shows the portion of the IMF that is approximated as continuous after sampling. The blue arrows represent individual, discrete high-mass stars tracked within the star particle. The difference in number of high-mass stars from the 5.5 stars expected above 8~M$_\ssun$ using a continuous IMF, as well as the difference in mass of high-mass stars from the expected 105~M$_\ssun$, is given for each realization. A stochastically populated IMF leads to large variation in mass and number of high-mass stars in each star particle.}
    \label{fig:stochdemo}
\end{figure*}

\begin{figure*}
    \centering
    \includegraphics[width=0.85\textwidth]{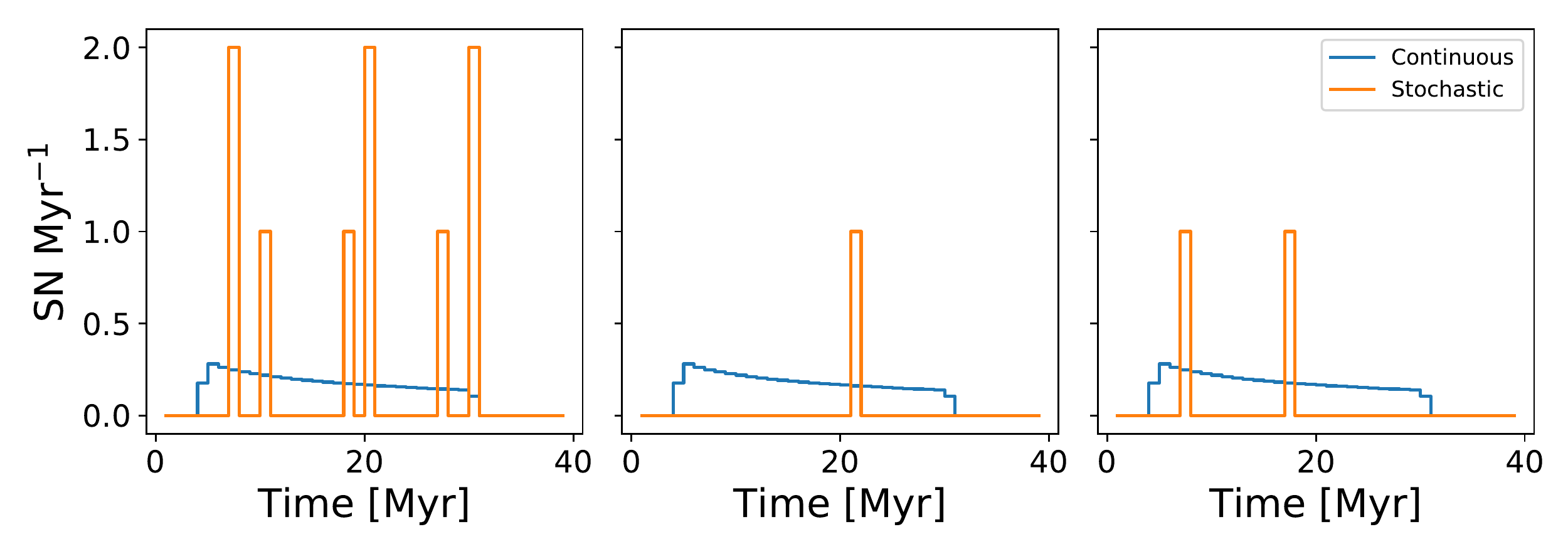}
    \caption{The supernova rate in 1~Myr intervals for the same three star partices as Fig.~\ref{fig:stochdemo}, with a stochastic IMF and a continuous IMF. The stochastically populated IMF leads to long periods of no feedback, with several intervals of stronger feedback. There are also different total numbers of supernovae and greatly varied timings. Though the middle star particle has three discretely tracked high-mass stars, only one star is below 40~M$_\ssun$ and therefore in the mass range of Type II supernovae.}
    \label{fig:sndemo}
\end{figure*}

An example of the methodology can be seen in Fig.~\ref{fig:stochdemo}, where we show three different realizations of a stochastically populated star particle with a total mass of 500 M$_\ssun$. In a population of this mass, we expect approximately 5.5 stars with 8 M$_{\ssun}<$~m$_\mathrm{star}$~$<100$ M$_\ssun$, and 105 M$_\ssun$ in mass for the same range. For each realization we show the change in number and mass of high-mass stars from these expected values, as well as the percent changes these correspond to. Since only about 20 per cent of the mass in these star particles is in the high-mass range, even relatively large deviations in the mass content of the discrete portion result in apparently small changes in the normalization of the low-mass portion. To see how the sampling affects the resulting stellar feedback, Fig.~\ref{fig:sndemo} shows the supernova rate in 1~Myr time steps for the same three star particles.

These three realizations demonstrate several important features introduced by a stochastically populated IMF. The first, and most fundamental, is that high-mass stars are discretized, and therefore Type II supernovae are discretized as well. We discuss in Section~\ref{sec:quant} how proper discretization of stars can \textit{only} be done with a comprehensive consideration of IMF sampling methodology. Another unsurprising change is that the actual number of high-mass stars within a star particle can vary greatly. For example, there are 8 stars with m$_\mathrm{star} > 8$ M$_\ssun$ in the left-most star particle of Fig.~\ref{fig:stochdemo}, and only two in the right-most star particle. Less obvious, the number and mass in high-mass stars is only loosely correlated. The middle star particle, for example, has fewer high-mass stars than the average, but has more mass in this range because the few high-mass stars that did form tended to be more massive. As seen in the middle panel of Fig.~\ref{fig:sndemo}, we also note that since Type II supernovae are restricted to stars with 8 M$_\ssun<$ m$_\mathrm{star}$ $<40$ M$_\ssun$, only one of the high-mass stars in this star particle would actually explode as a supernova.

\subsection{Existing discretization methods}\label{sec:quant}

Currently, there is one dominant method of discretizing supernova feedback in cosmological simulations \citep[e.g.][]{Stinson2010, Hopkins2014, Smith2018, Rosdahl2018, su2018}, which we will henceforth refer to as ``quantized feedback.'' While this is not the only method used in the simulation community, it seems to be the prevailing method among high-resolution cosmological simulations, and so we focus on it here. We emphasize that the issues raised in the following discussion are true for any method that does not sample the distribution of stars at birth, but rather calculates supernova explosions ``on the fly'', while leaving the remainder of the IMF unchanged.

In quantized feedback, the number of supernovae in a given time step is drawn from either a binomial distribution or Poisson distribution\footnote{A binomial distribution with many trials and a small probability of success converges to a Poisson distribution, so these two formulations are approximately equivalent.}. The supernova mean rate may be taken from rate tables \citep[as in][]{Hopkins2014} or from the expectation number of supernovae in that time step (e.g. \citealt{Stinson2010} or the RIMFS method of \citealt{Revaz2016}). While this method guarantees that only integer numbers of stars explode in a given time step, we note that at high resolution there are several internal inconsistencies upon closer inspection. The importance of these inconsistencies on simulation results depends on the size of the ensemble of star particles considered. Here, we start with the case of a single star particle.

First, this method cannot guarantee that the mass used in calculating feedback is the same as the dynamical mass of the particle. The mass of stars below the quantized regime is fixed, while the mass of stars within the quantized regime can vary by more than a factor of 2. In terms of the number of stars within the star particle, a fixed number of stars (those below the quantization limit) are added to a Poisson distribution of stars (those that are quantized). This combination will rarely yield the assumed initial particle mass. For smaller star particles, this effect will be more severe. The star particle mass and the mass removed from any individual parent gas particle will therefore be inconsistent. If, on the other hand, mass conservation is enforced (for example, by adding back missing mass or subtracting excess mass in the low-mass end after supernovae explode), then this method will require the mass in low-mass stars to be amended in real time in an unphysical way, leading to more internal inconsistencies. Additionally, imposing a limit (minimum) on the number of supernovae that can explode will artificially concentrate supernovae to go off in early (late) times.

\begin{figure}
    \centering
    \includegraphics[width=0.45\textwidth]{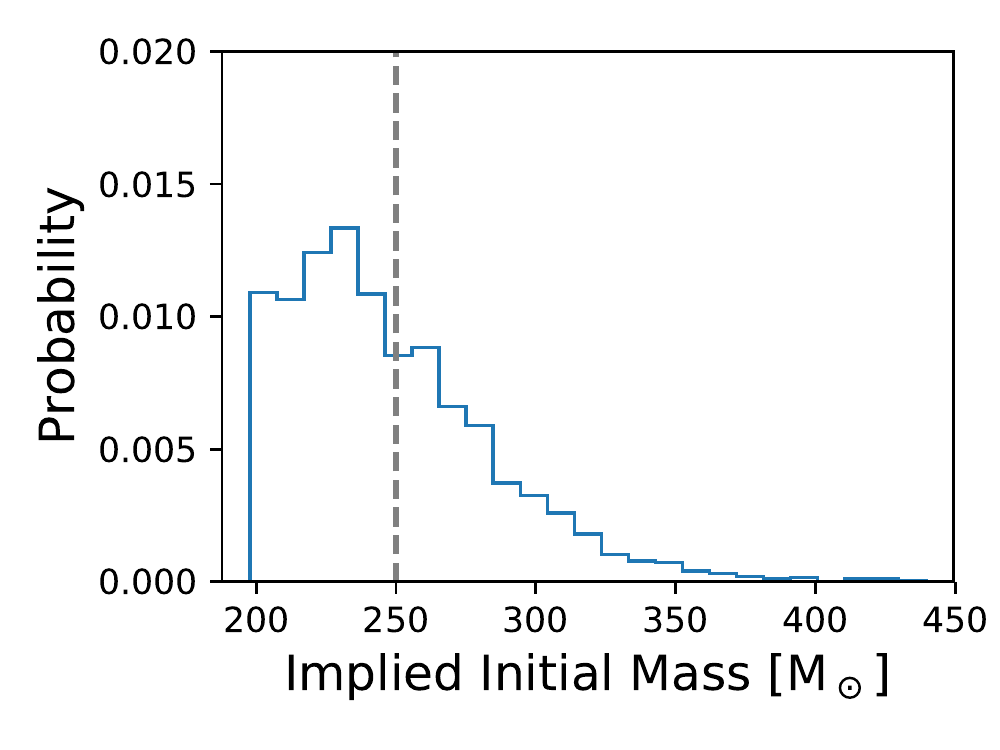}
    \caption{A Monte Carlo simulation of quantized feedback, in which supernovae explode each time step according to a binomial distribution given by their expected number. Thus, the mass in stars less massive than Type II progenitors is fixed while the mass in high-mass stars can vary widely. The probability distribution of the resultant total initial mass of the star particle is shown. The vertical dashed line shows the assumed initial mass. Quantized feedback can result in almost 40 per cent difference between the assumed initial mass of the star particle and the mass implied by the number and mass of supernova explosions. To enforce mass conservation, the number and mass in low-mass stars would have to be constantly updated in real time as supernovae explode, leading to further inconsistencies.
    }
    \label{fig:quant}
\end{figure}

To see this in more detail, we consider many trials of a 250 M$_\ssun$ star particle with a \citet{Kroupa2001} IMF, and demonstrate that without an a priori knowledge of the stellar mass distribution, the stated star particle mass is inconsistent with the initial mass implied by the number of supernovae explosions. We consider a slightly simplified version of what occurs in simulations, assuming for each trial that we begin with the same metallicity. The mass below 8 M$_\ssun$ in a star particle of this mass and IMF is $198$ M$_\ssun$, with the remainder falling in the high-mass portion of the IMF. For each particle, we iterate forward in time steps of $10^5$ years, and at each time step determine whether a supernova explodes by the method of \citet{Stinson2010} (i.e. drawing from a binomial distribution). To make this comparison more applicable to other simulators without upper limits on core collapse supernovae, we do not place a Type II upper limit of 40~M$_\ssun$. If a supernova explodes, we add the initial mass of the star that exploded to the initial low-mass total of the star particle. Since the range outside of Type II supernovae is not quantized, its mass is unaffected by this procedure. The results of this process are shown in Fig.~\ref{fig:quant}. While we use the specific methodology of \citet{Stinson2010}, the results of Fig.~\ref{fig:quant} are broadly true for any scheme that uses a binomial or Poisson distribution to determine whether a supernova explodes in real time at each time step.

A second drawback of this method, similar to the first, is that other forms of feedback are incorrectly calculated for every individual star particle. Much as mass is not conserved because the number and mass of high-mass stars is not known a priori, neither is any form of feedback that relies on high-mass stars. Photo-ionizing radiation and photo-electric feedback in simulations are calculated from the distribution of stars within a population, often relying on stellar population synthesis codes \citep[e.g.][]{Hopkins2012, Christensen2012, Agertz2013, Stinson2013, Rosdahl2013, Ceverino2014}. If the actual distribution is not known ahead of time (and, as shown previously, is almost certainly different from the assumed distribution), then the photon estimates will be inconsistent. Since various feedback effects add non-linearly \citep{Hopkins2014}, any simulation that includes different feedback mechanisms must ensure that all are consistent. The only way to ensure consistency is to sample the IMF at the formation time of the particle.

To see how the results diverge between quantized feedback and a stochastic IMF, we run an ensemble of dwarf galaxies using the quantized feedback of \citet{Stinson2010}. We compare the results in Section~\ref{sec:quantresults}.

We note one more advantage of the stochastic IMF over quantized feedback: any other sampling method can be easily incorporated. As discussed above, there is still ongoing debate over the fundamental way in which high-mass stars form, and therefore the proper way to sample from the IMF. If we wanted to use, for example, the sorted sampling of \citet{Weidner2006}, we could do so provided we implement a treatment for clusters. Choosing sampling methods is impossible in methods such as quantized feedback where the mass distribution is not known from birth.

\section{Results}\label{sec:results}

\subsection{Cosmological star formation}\label{sec:sf}

For this paper, we restrict our attention to galaxies residing in well-resolved halos with M$_\mathrm{vir}>10^7~\mathrm{M}_\ssun$. We focus exclusively on galaxies that exist in both runs, to ensure our results are converged. To avoid Poisson noise in star formation between matching halos, we only consider galaxies where at least one of the two runs formed at least four star particles. At $z=6$, Elena forms 86 galaxies satisfying the above criteria that are either in the field or satellites of the main halo.

The most direct way to measure the impact of any prescription is through its effect on the stellar mass of galaxies. There is active research in the literature regarding the abundance of dwarf galaxies, which will soon be discovered in unprecedented numbers with the Large Synoptic Survey Telescope. There is also growing consensus that there is large scatter in the stellar mass-halo mass relationship at the low-mass end \citep{Lin2016, Garrison-Kimmel2017, Munshi2017, Kulier2019}. While we do not expect a stochastic IMF to significantly alter these relationships in massive galaxies, we do expect a resulting change in feedback to affect dwarf galaxies, particularly those small enough such that their IMF is not converged.

\begin{figure*}
    \centering
    \includegraphics[width=0.8\textwidth]{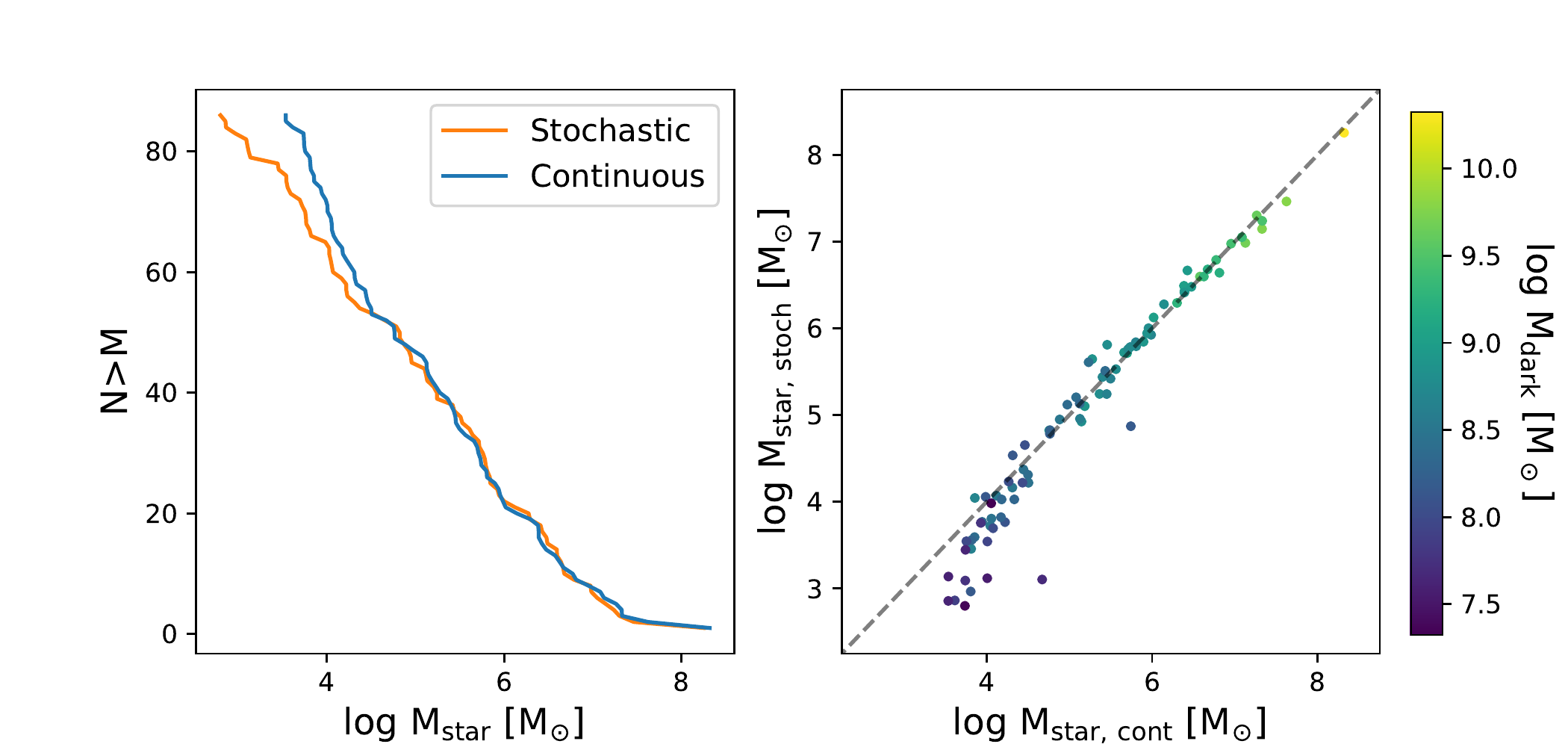}
    \caption{\textit{Left}: The cumulative stellar mass function of galaxies in our sample, for runs with a stochastic and continuous IMF. While converged above ${\sim}10^5$~M$_\ssun$, the run with a stochastic IMF is shifted towards smaller stellar masses below this mass. \textit{Right}: Stellar masses for galaxies residing in matching halos between the two simulations. Points are colored according to the mass of the dark matter halo hosting the galaxy. The dashed grey line shows equal masses between runs. The smallest galaxies see a systematic reduction in stellar mass with a stochastic IMF compared to the run without a stochastic IMF.}
    \label{fig:smf}
\end{figure*}

The left panel of Fig.~\ref{fig:smf} shows the cumulative stellar mass functions at $z=6$ of the cosmological runs. The runs are identical other than the treatment of the IMF. Galaxies with M$_\mathrm{star}\gtrsim10^5~\mathrm{M}_\ssun$ are fully converged, with identical stellar mass functions above this point. Below this mass, however, there is a clear shift towards smaller stellar masses in the run with a stochastic IMF.

To further quantify the differences in stellar mass, the right panel of Fig.~\ref{fig:smf} shows the stellar masses of all galaxies in the left panel, with corresponding galaxies matched between the two runs. The galaxies are colored according to the mass of the dark matter halo in which they reside\footnote{More precisely, they are colored by the mean of matching halo masses in the two runs; however, the dark matter halo masses of the stochastic and non-stochastic runs are typically not more than 5 per cent different.}. The figure shows that in the smallest galaxies, the stellar mass is generally lower with the stochastic prescription than with the continuous prescription. In the low-mass range, galaxies with a stochastic IMF see a reduction in stellar mass of up to an order of magnitude compared to the equivalent galaxy with a continuous IMF.

\begin{figure}
    \centering
    \includegraphics[width=0.45\textwidth]{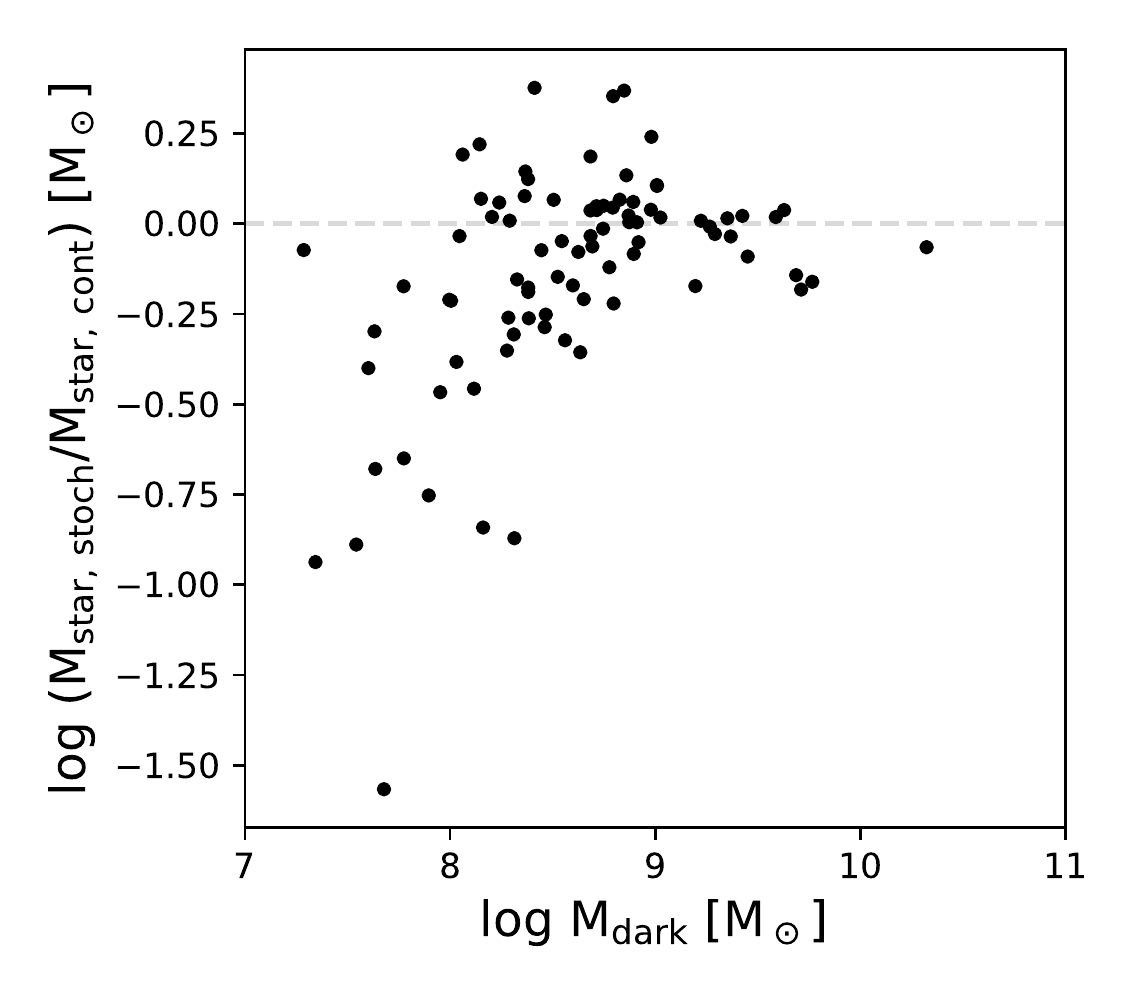}
    \caption{Difference in stellar mass as a function of dark matter halo mass for matched galaxies across runs with and without a stochastic IMF. The dashed grey line indicates equal masses between runs. Lower halo masses have increased scatter, and starting at M$_\mathrm{dark}\sim~10^{8.5}$M$_\ssun$, galaxies in the run with a stochastic IMF are generally lower than with a continuous IMF.}
    \label{fig:dmstar}
\end{figure}

The trend of star formation suppression with a stochastic IMF becomes clearer if we instead arrange the stellar masses according to their dark matter masses. Fig.~\ref{fig:dmstar} shows the change in stellar mass as a function of halo mass for both sets of galaxies. Below a halo mass of ${\sim}10^{8.5}$~M$_\ssun$, galaxies formed with a stochastic IMF tend to have their star formation suppressed compared to a continuous IMF. Above this mass there is little change. This trend appears in addition to the generally greater scatter at lower halo mass which results from the breakdown of halo self-regulation. This dependence clarifies one of the outlier results of Fig.~\ref{fig:smf}, where a somewhat more massive galaxy (${\sim}10^6$~M$_\ssun$ in the run with a continuous IMF) still sees a reduction of an order of magnitude in stellar mass in the run with a stochastic IMF. The cause for this is that while its stellar mass is greater, its dark matter halo is more typical of the type that hosts ultra-faint galaxies.

\subsection{Bursty Feedback}

\subsubsection{Supernova timing}

To explain the reduced star formation with a stochastically sampled IMF as compared to the continuous IMF, a simple guess would be that since we now allow the total number of supernovae per star particle to vary, we now have varying total levels of feedback. At low SFR, however, a stochastically sampled IMF is expected to under-fill the high-mass end of the IMF, which would lead to fewer supernovae and less energy in feedback. If total supernova energy were the dominant factor, then, we would expect to see higher stellar masses in the run with a stochastic IMF, which is the opposite of the results shown in Fig.~\ref{fig:smf}. In fact, among galaxies with suppressed star formation, roughly equal numbers have above and below average supernova total feedback.

\begin{figure}
    \centering
    \includegraphics[width=0.45\textwidth]{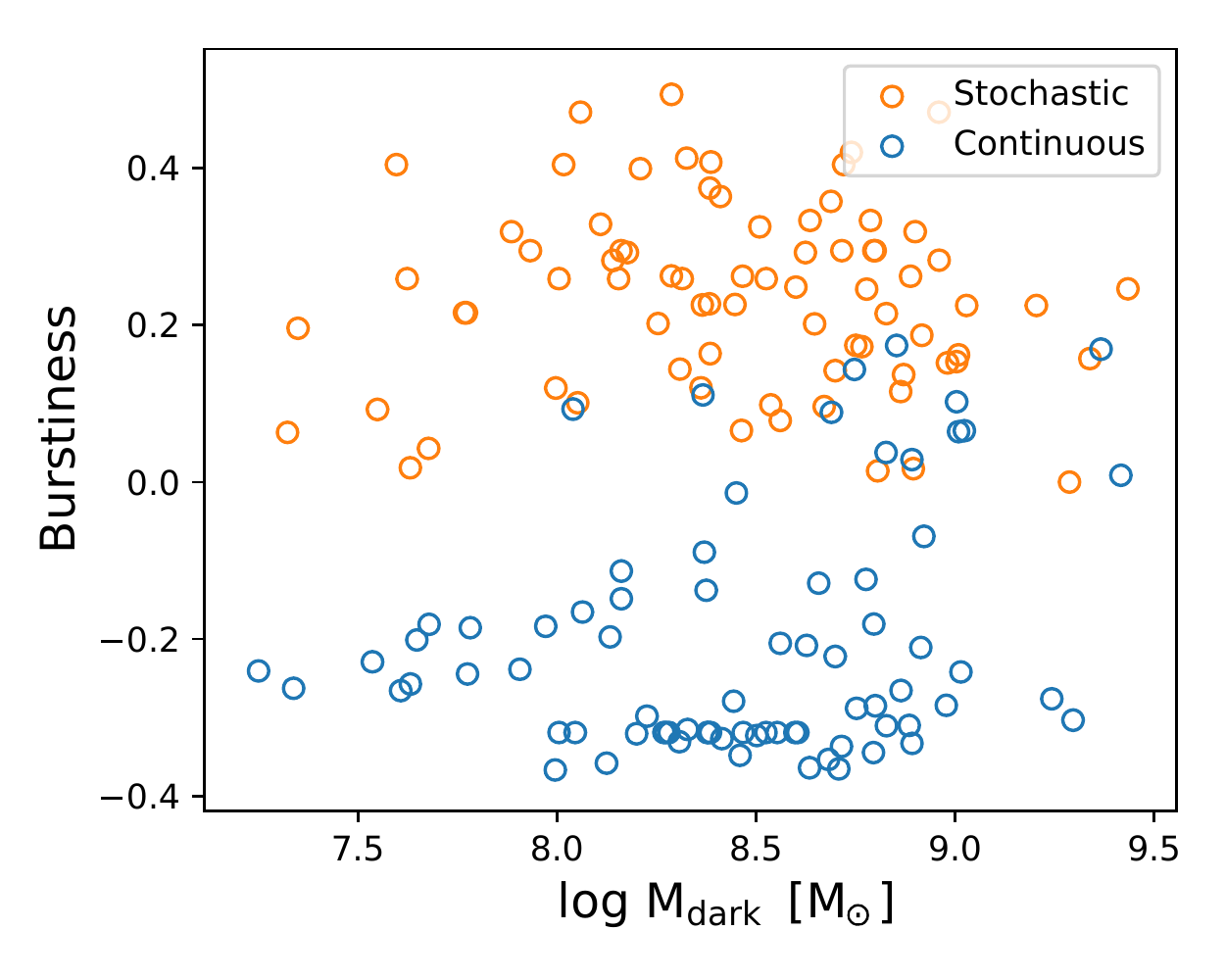}
    \caption{Burstiness of the supernova rate as a function of dark matter halo mass for runs with and without a stochastic IMF. Burstiness is calculated according to equation \ref{eq:burstiness} using 1~Myr intervals during the first 30~Myr of star formation. Independent of galaxy properties, the run with a stochastic IMF leads to burstier supernova feedback, which leads to the suppression of star formation in dwarf galaxies that are too low-mass to self-regulate.}
    \label{fig:burst}
\end{figure}

More important than the absolute number of supernovae, then, is the timing of the explosions. As is clear in Fig.~\ref{fig:sndemo}, the supernova feedback with a stochastic IMF becomes much more temporally clustered. To see this in more detail, we can define a burstiness parameter (\citealt{Goh2008}, and similar to equation~1 of \citealt{Mistani2016}) as
\begin{equation}
  B=\frac{\sigma/\mu-1}{\sigma/\mu+1},\label{eq:burstiness}
\end{equation}
  
where $\sigma$ is the standard deviation of the supernova rate, and $\mu$ is the mean supernova rate. Using this definition, the burstiness ranges from $-1$ to $1$; a uniform distribution has a burstiness $B=-1$, an exponential distribution has a burstiness $B=0$, and the burstiness approaches 1 as $\sigma/\mu\to\infty$. We calculate the rate in 1~Myr intervals\footnote{The burstiness parameter will be somewhat dependent on the binning chosen for the supernova rate, but we confirmed that the increased burstiness of the stochastic IMF over the continuous IMF is independent of bin size. We choose binning in 1~Myr intervals to be consistent with the timing of feedback in our simulations.} for the first 30~Myr of every galaxy's star formation (coinciding roughly with the longest-lived supernova from the first star particle to form in the galaxy). The results are plotted in Fig.~\ref{fig:burst}.

As seen in the figure, the supernova rate is significantly burstier in runs with a stochastic IMF than runs with a continuous IMF. Crucially, the increase in burstiness applies to all galaxies, not only galaxies with low SFR. Since feedback with a stochastic IMF is universally more effective, it may seem surprising that we only see an impact in small galaxies. As was shown in Fig.~\ref{fig:dmstar}, this stems from the stronger dependence on halo mass than stellar mass. The bursty supernova feedback leads to more effective heating of gas, as will be shown explicitly in the next section. However, only in small halos less able to self-regulate does this more effective feedback completely shut off future star formation. The deeper potential wells of the higher mass halos minimize the effects of burstier feedback.

\subsubsection{Isolated runs}\label{sec:isolated}

The same simulation run multiple times can have differing galaxy properties, owing to stochastic variations in numerical codes \citep{Keller2019}. To further investigate the results of the previous section and to ensure our results are independent from the intrinsic scatter in galaxy stellar mass, we consider an isolated simulation of a $10^9$~M$_\ssun$ halo, with initial star particle masses of 420~M$_\ssun$.

\begin{figure}
    \centering
    \includegraphics[width=0.45\textwidth]{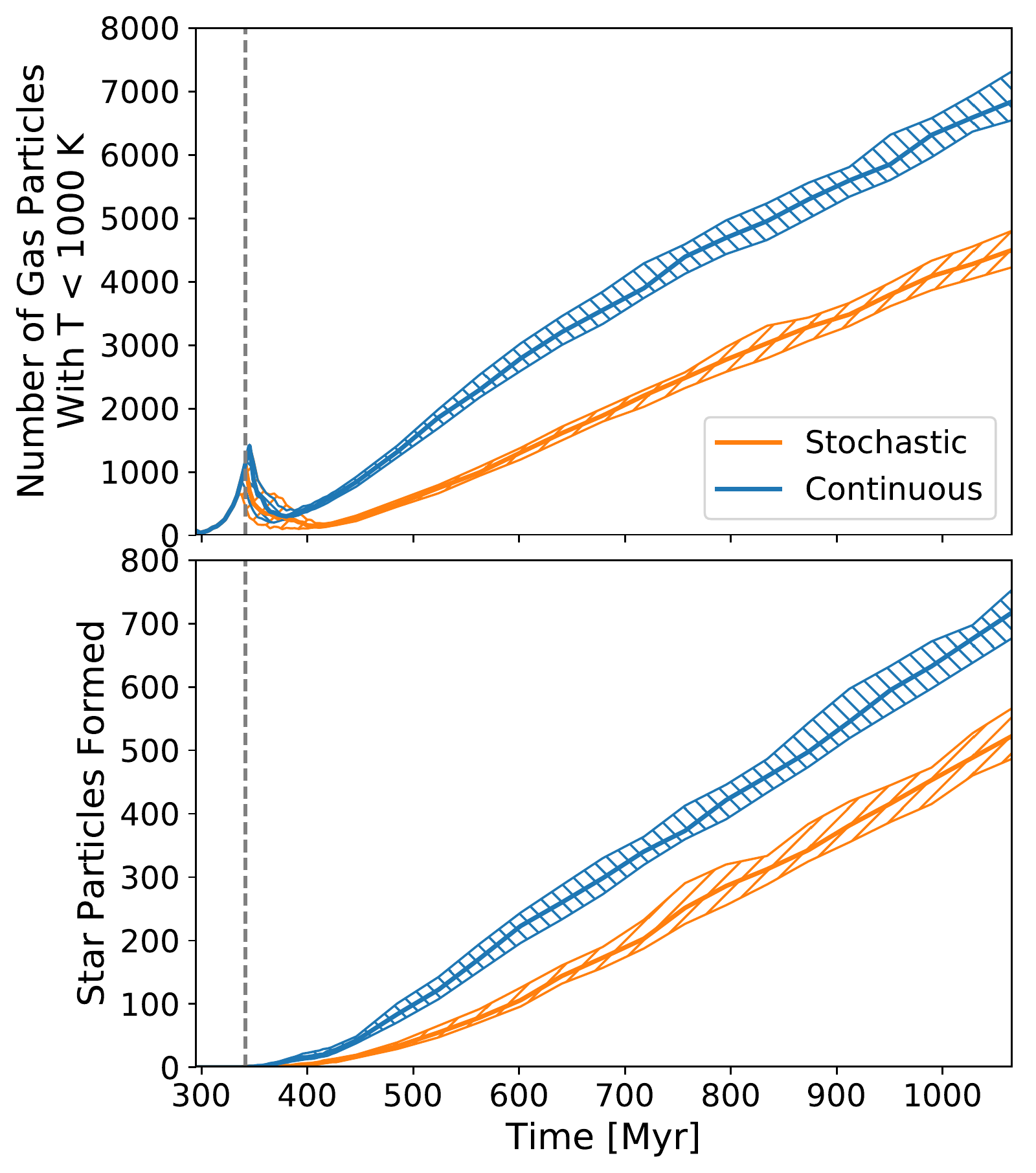}
    \caption{Star formation in the isolated dwarf galaxy runs, with and without a stochastic IMF. Each IMF treatment was run 50 times. Central lines show the medians, and bands represent the interquartile range. The grey line represents the median time of the first star formation. \textit{Top}: The available cold gas as a function of time. With a stochastic IMF, about half as much gas is available compared to runs with a continuous IMF. \textit{Bottom}: Cumulative star formation as a function of time. With a stochatic IMF, star formation is suppressed compared to runs with a continuous IMF, resulting from supernova feedback more effectively heating surrounding gas (see top panel).}
    \label{fig:numtemp}
\end{figure}

To quantify the significance of differences between IMF treatments, we simulate the isolated halo 50 times with a stochastic IMF and 50 times without. The results are shown in Fig.~\ref{fig:numtemp}, where we focus on the first Gyr. The bottom panel of the figure shows the cumulative number of star particles that have formed as a function of time, displaying both the median of all runs and the interquartile range. Clearly, the majority of the time, the run with a stochastic IMF forms fewer stars throughout the entire duration of the simulation.

To see why star formation is suppressed, the top panel of Fig.~\ref{fig:numtemp} shows the number of gas particles with a temperature below 1000~K. While the conditions for star formation are based on H$_2$ abundance \citep{Christensen2012}, this serves as a proxy for the number of gas particles that could potentially form stars. We see that with a stochastic IMF, the gas is more effectively heated---at any given time, up to half as many gas particles have cooled to below 1000~K, in line with the reduction in star formation. We have verified that in our cosmological simulation, the galaxies tend to have similar gas masses, with no systematic change as with stellar mass. This indicates that more effective gas heating dominates over more efficient gas expulsion.

In a cosmological setting, these effects can be significantly exaggerated. In the cosmological runs compared to the isolated runs, the differences between the stellar masses of galaxies can increase from a factor of ${\sim}1.5$ to ${\sim}10$, as in Fig.~\ref{fig:smf}.

\subsection{Metallicity}

Beyond energy deposition, high-mass stars return processed material to the ISM. As previously emphasized, a stochastic IMF significantly alters the distribution of high-mass stars, including their masses and numbers. Since the metal production depends non-linearly on the mass of the exploding star \citep{Raiteri1996}, one might expect greater variation in the chemical enrichment of galaxies with a stochastic IMF. 

\begin{figure}
    \centering
    \includegraphics[width=0.45\textwidth]{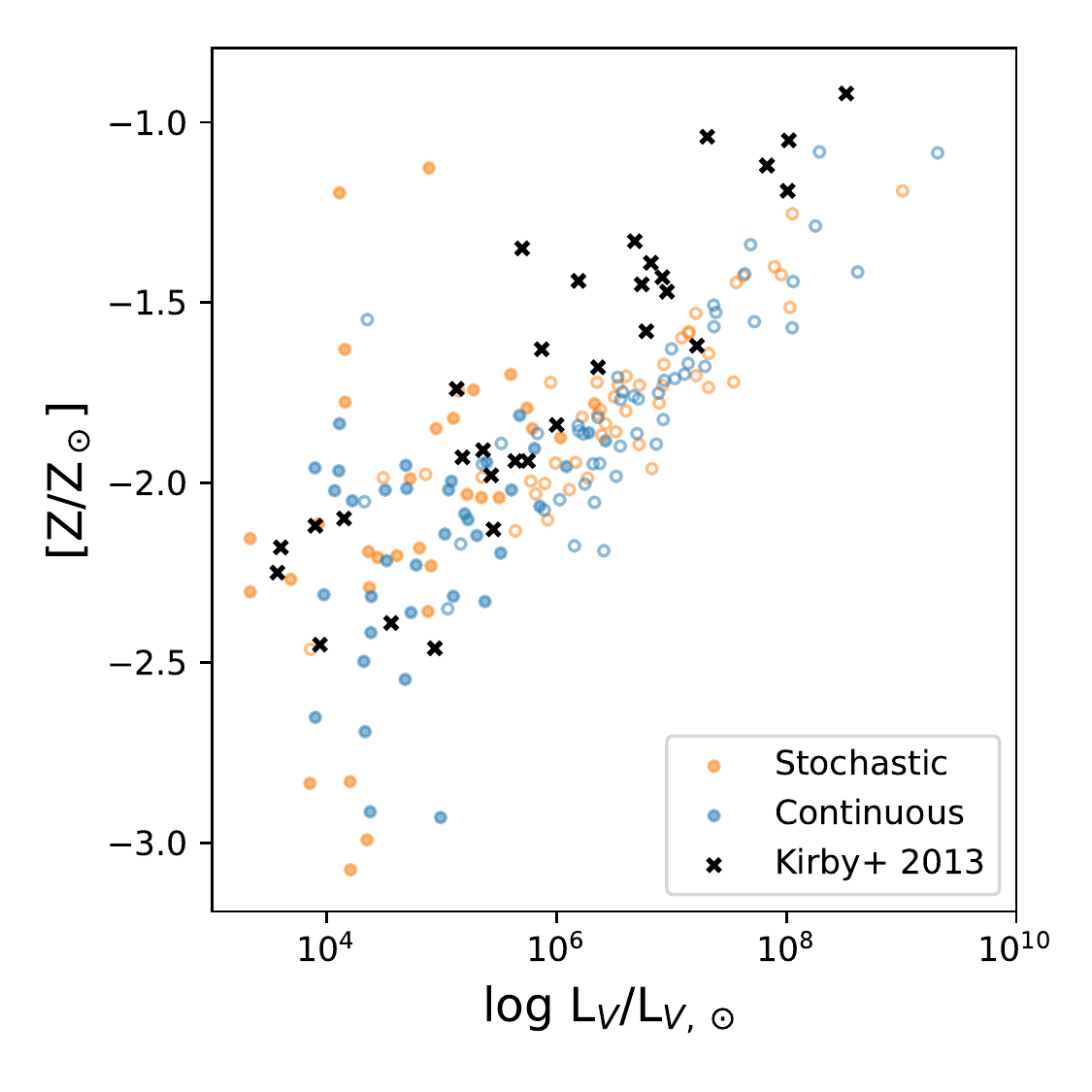}
    \caption{Mean stellar metallicity as a function of luminosity for galaxies in our sample, with and without a stochastic IMF. Filled circles show galaxies residing in dark matter halos below $10^{8.5}$~M$_\ssun$, approximately corresponding to galaxies we expect to be quenched by reionization \citep{Tollerud2018}. Since the sample is at $z=6$, only low-mass galaxies that have stopped forming stars match the $z=0$ data. More massive galaxies are expected to increase in metallicty with time. There are no clear differences between the IMF treatements, though there may be slightly greater scatter in metallicity with a stochastic IMF.}
    \label{fig:metals}
\end{figure}

Fig.~\ref{fig:metals} shows the luminosity-metallicity relationship of all galaxies in the sample, along with data from \citet{Kirby2013}. To calculate the galaxy metallicities, we apply a floor for individual star particles of $Z>10^{-5}$. Luminosities are calculated using \textsc{PARSEC}\footnote{http://stev.oapd.inaf.it/cgi-bin/cmd} isochrones \citep{Bressan2012}. We note that our sample is at $z=6$; for faint galaxies that have likely stopped forming stars indefinitely, we match the data well, though our scatter below $L_V\sim10^5$~$L_{V,\ssun}$ is higher than the observations. Some of the excess scatter is likely reduced due to mergers at later times, when some of the faint galaxies with extremely low metallicities are incorporated into larger, higher metallicity galaxies. For larger galaxies that lie below the $z=0$ observations, ongoing star formation will increase the metallicities with time. A forthcoming paper (Munshi~et~al.~in~prep) will discuss these relationships and demonstrate consistency with the data in the present day.

Surprisingly, runs with the stochastic and continuous IMF are consistent at all luminosities, though the scatter may increase for faint galaxies with a stochastic IMF. It is likely that metal diffusion in the ISM \citep{Shen2010} quickly obscures any systematic differences in future generations of stars that form, consistent with the findings of \citet{Revaz2016}, who found that introducing metal diffusion in dwarf galaxies reduced scatter introduced by IMF sampling effects. The overall impression is that the stochastic IMF has little impact on the chemical evolution of galaxies.

\begin{figure}
    \centering
    \includegraphics[width=0.5\textwidth]{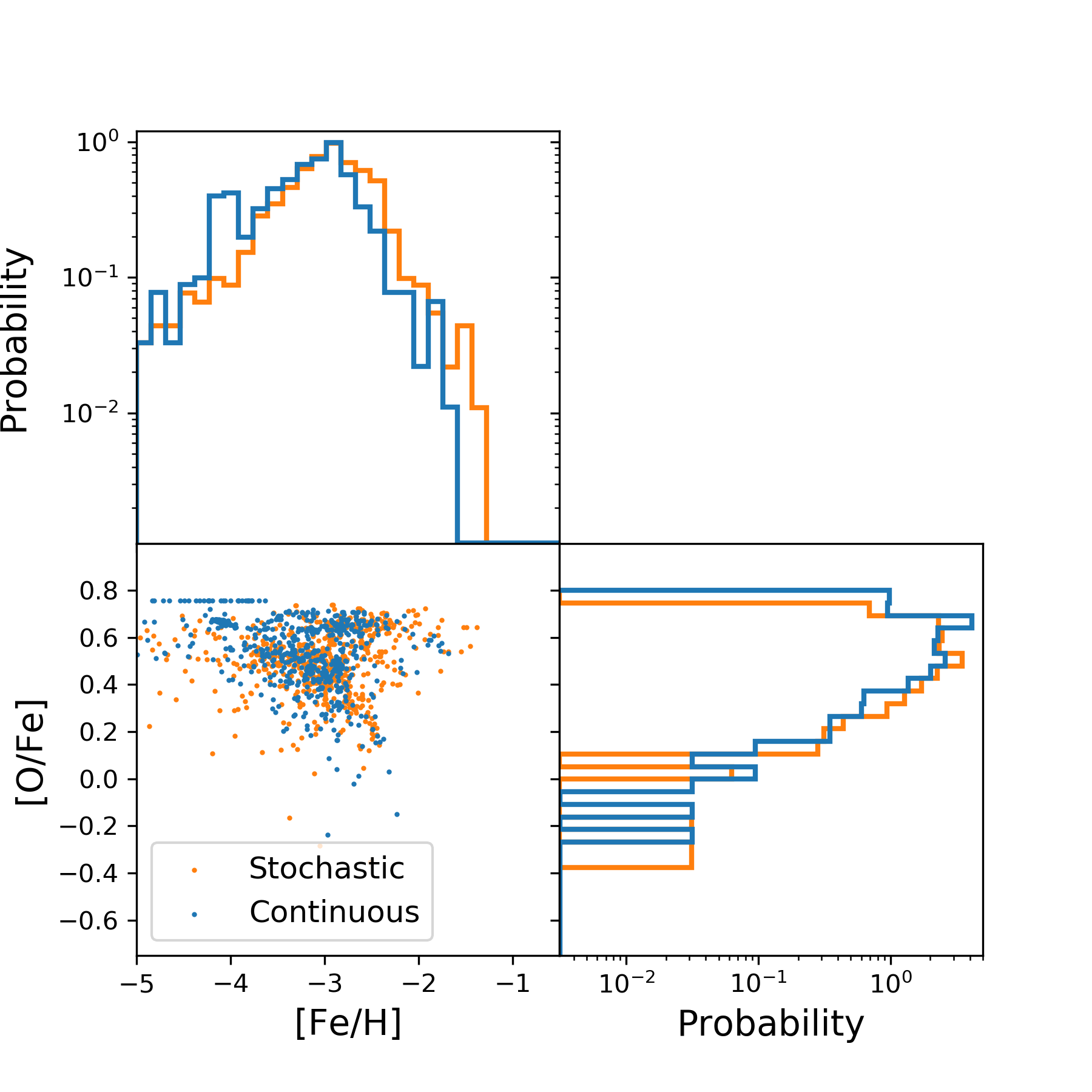}
    \caption{Chemical abundances of all star particles in galaxies with $z=6$ luminosities $L_V\leq10^{5.5}$~$L_{V,\ssun}$ and M$_\mathrm{dark}\leq10^{8.5}$~M$_\ssun$, corresponding roughly to $z=0$ UFDs. The abundance ratios between the two models are largely consistent, likely as a result of metal diffusion in the ISM, which reduces scatter introduced by IMF sampling effects \citep{Revaz2016}. The high number of stars at $\mathrm{[O/Fe]}\sim0.75$ using a continuous IMF result from the stars formed from the metal content of the very first time step in which supernovae explode. There may be a slight shift to higher [Fe/H] with a stochastic IMF, but it is well below observational precision.}
    \label{fig:alphairon}
\end{figure}

We also investigate whether the stellar chemical abundances change with a stochastically populated IMF. Figure~\ref{fig:alphairon} shows the abundance ratios for galaxies with ($z=6$) luminosities $L_V\leq10^{5.5}$~$L_{V,\ssun}$ and dark matter masses below $10^{8.5}$~M$_\ssun$, which after passive stellar evolution would correspond roughly to today's UFDs. As above, there is almost no difference between the models; the [O/Fe] vs [Fe/H] distributions overlap completely. With the continuous IMF there is a small overabundance of stars at $\mathrm{[O/Fe]}\sim0.75$, which corresponds to the maximum abundance ratio possible that results from the very first time step in which supernovae explode when integrating along the continuous IMF. Additionally, there is a mild indication that stars with a stochastic IMF are shifted to slightly higher [Fe/H]. This shift makes sense given that requiring whole supernovae to explode means that more processed material is ejected at once into the ISM. On the whole, however, it seems that even when looking at the chemical composition of stars there is little to distinguish the models, especially given observational uncertainties.

\subsection{Stochastic vs quantized feedback}\label{sec:quantresults}

\begin{figure}
    \centering
    \includegraphics[width=0.45\textwidth]{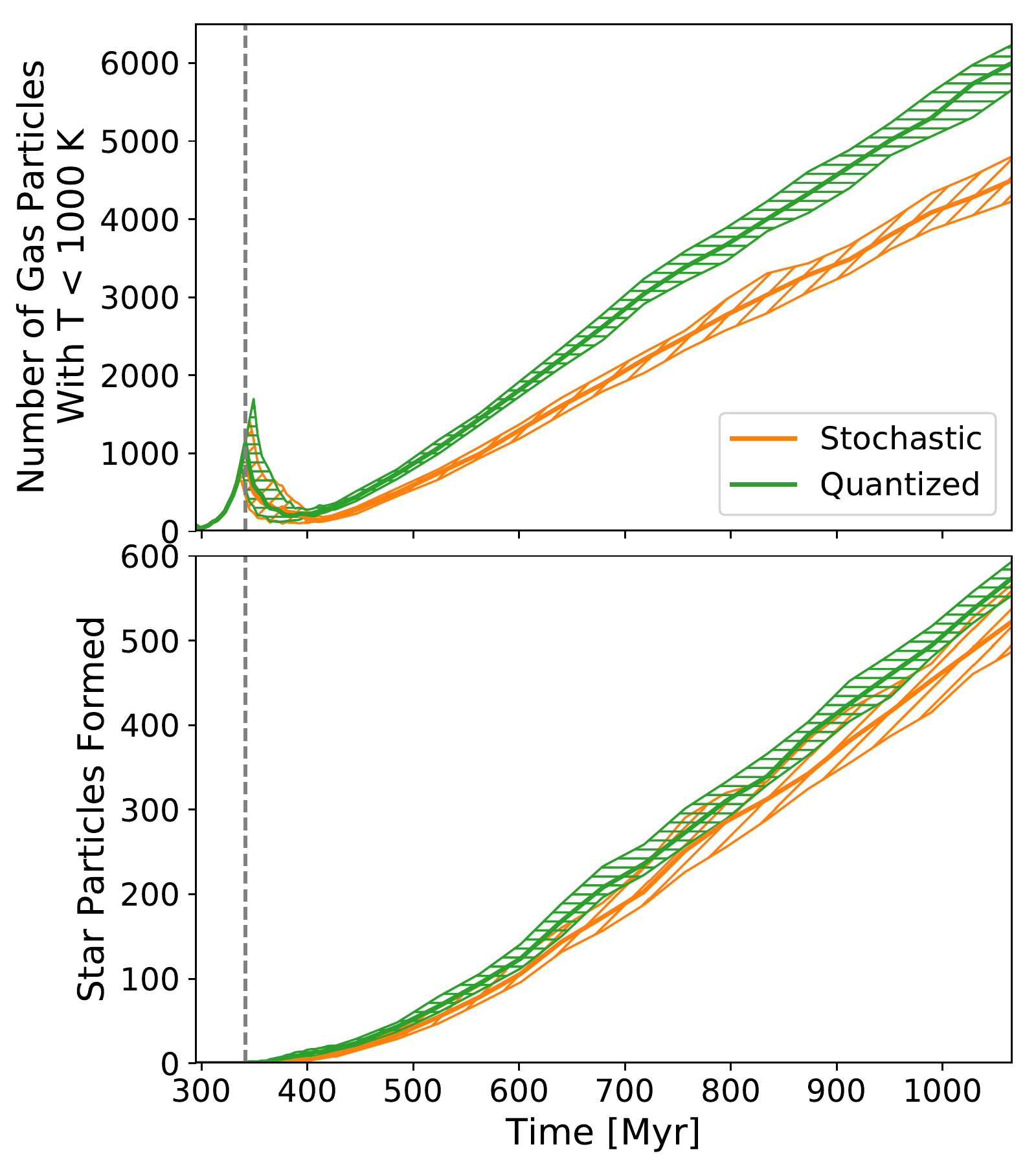}
    \caption{Star formation in the isolated dwarf galaxy runs, comparing runs with a stochastic IMF vs runs with quantized feedback. Each IMF treatment was run 50 times. Central lines show the medians, and bands represent the interquartile range. The grey line represents the median time of the first star formation. \textit{Top}: The available cold gas as a function of time. Quantized feedback leaves ${\sim}30$ per cent more cold gas available to form stars compared to runs with a stochastic IMF. \textit{Bottom}: Cumulative star formation as a function of time. Quantized feedback results in ${\sim}$10 per cent more star formation than a stochastic IMF.}
    \label{fig:numtempquant}
\end{figure}

As discussed in Section~\ref{sec:quant}, quantized feedback is an existing method that discretizes supernovae. To test how it compares to a stochastic IMF, we ran the isolated dwarf galaxy 50 times with quantized Type II supernovae (but otherwise the same feedback implementation). Fig.~\ref{fig:numtempquant} shows the quantity of cold gas and cumulative star formation as a function of time for the two IMF treatments. Interestingly, there is consistently ${\sim}30$ per cent more available cold gas and ${\sim}10$ per cent more star formation with the quantized feedback as compared to the stochastic IMF.

Stellar feedback, therefore, is less effective with quantized feedback than with a fully self-consistent stochastic IMF. There are multiple differences that could contribute to this behavior. First, as discussed in Section~\ref{sec:quant}, the LW photon production with quantized feedback cannot accurately reflect the internal distribution of high-mass stars. Different feedback mechanisms interact non-linearly \citep{Hopkins2014}, such that one can reinforce the effect of the others. The stochastic IMF ensures that LW radiation and supernovae come from the same stars, which may result in stronger disruption of gas. Additionally, since UV luminosities depend highly non-linearly on stellar mass, estimating the ionizing photon counts from fractions of massive stars may result in less initial LW radiation than for individual (whole) massive stars, though averaged over many particles the LW outputs converge. Another difference may come from the timing and distribution of supernovae; if the stochastic IMF leads to more temporally clustered explosions, then the cumulative heating of gas may be more effective. Monte Carlo simulations of a stochastically populated IMF show that approximately 25 per cent of 420~M$_\ssun$ star particles with feedback calculated every 1~Myr will have a single time step with multiple supernovae, as opposed to none using quantized feedback. Thus, some of the difference could result from our use of a binomial rather than Poisson distribution in calculating the quantized feedback, and we caution that a different implementation may yield closer results. We note that the average total number of supernovae per star particle is the same between the prescriptions.

We caution that these results come from an isolated dwarf galaxy. In a cosmological context, the presence of tidal interactions, ionizing radiation, and gas outflows may increase the differences in these results. In particular, the presence of a larger cold gas reservoir when using quantized feedback could lead to more extended star formation histories in ultra-faint dwarf galaxies during and after the epoch of reionization.

In our case, because the primary source of energetic feedback comes from Type II supernovae, the differences between a stochastic IMF and quantized feedback are conspicuous but limited. This demonstrates the general robustness of the supernova feedback physics implemented in our simulations. Some simulations incorporate significant quantities of radiative feedback from high-mass stars, beyond merely a Lyman-Werner prescription for H$_2$ destruction. While capturing more subgrid processes, they may also suffer from an even greater internal inconsistency in the use of quantized feedback, due to the non-linear nature of different feedback mechanisms, which may result in even greater differences between quantized feedback and a stochastically sampled IMF.

\section{Implications for future predictions}\label{sec:discussion}

\subsection{Star formation quenching}\label{sec:sfquenching}

The most significant difference between the stochastic IMF and the continuous IMF is the lower stellar mass of a large fraction of dwarf galaxies. This is not unexpected, and is similar to what was found by \citet{su2018}. The effect we find, however, is less extreme, owing mostly to the differences in our feedback implementations. Our feedback is calculated in 1~Myr intervals, compared to the smaller time scales in the \textsc{fire} simulations, which can be shorter than $10^4$~yr. Thus, continuous injection of supernova energy results in relatively large per-step feedback in our simulations as compared to continuous injection in \textsc{fire}. Additionally, in \textsc{ChaNGa} instantaneous energy deposition is incorporated as a 1~Myr heating rate to avoid numerical instabilities (see footnote~63 in \citealt{Kim2016}), meaning that even with a stochastic IMF supernova energy is effectively continuously injected when considering sufficiently small time intervals.

The cause of this suppression in star formation is clear from its effects on the gas in the isolated dwarf runs: supernova feedback with a stochastic IMF is more effectively preventing gas from cooling. In the first Gyr of star formation, there can be more than a factor of two difference in the amount of gas available to form stars, with a corresponding difference in the amount of star formation. The reason for the suppression of star formation is that supernova energy is deposited in a shorter time frame; this is not only because stars are discrete, but because stochasticity in the stellar masses allows supernovae to cluster temporally, as was shown in Fig.~\ref{fig:sndemo}.

We also find a strong halo mass dependence in our results, as was shown in Fig.~\ref{fig:dmstar}. Owing to our large sample, we are able to explore these trends. The continuous IMF treatment yields largely the same results for galaxies residing in halos more massive than ${\sim}10^{8.5}$~M$_\ssun$ at $z=6$. These more massive halos are able to self-regulate even with the stronger supernova feedback. Smaller halos, however, have too small a potential well to prevent supernovae from driving out gas from star forming regions. This is also approximately the (high-redshift) mass scale where reionization is thought to suppress star formation \citep{Quinn1996, Thoul1996, Barkana1999, Bullock2000, Gnedin2000, Okamoto2008, Tollerud2018}; most galaxies in halos of these masses will quench, if not from supernova feedback, then from the combined effects of feedback and reionization \citep{Benson2002, Somerville2002, Hoeft2006, Nickerson2011}. For a given reionization model, our results show that a stochastic IMF suppresses star formation even further in low mass halos.

\subsection{Reionization}

Though in some cases galaxies quenched at high redshift can restart star formation in later times \citep{Wright2019}, even a temporary suppression of star formation would have significant implications for the epoch of reionization. Early galaxies are thought to be the primary source of ionizing radiation \citep{Stark2016}, with significant contributions from dwarf galaxies. In fact, though not directly observable, inferences from the Local Group imply that very small dwarf galaxies (as faint as M$_V\sim-3$) may have contributed to reionization \citep{Weisz2017}. What is still unknown, however, is the fraction of ionizing radiation they provided, and similarly, their luminosity function at high redshifts. There have been many simulation predictions of the reionization era, reaching down to dwarf galaxy scales \citep[e.g.][]{O'Shea2015, Finlator2016, Gnedin2016, Ocvirk2016, Xu2016, Anderson2017, Ma2018}, and simulations continue to push to higher resolution.

When properly accounting for IMF sampling effects, we have seen that star formation is often suppressed earlier, indicating that by $z\sim6$, the faintest galaxies will constitute a reduced fraction of the ionizing photon budget. This may be countered by higher escape fractions resulting from hotter bubbles of gas around these small galaxies. Detailed explorations of the implications of IMF sampling on reionization will be pursued in future work, but what is already clear is that accurate predictions will require a stochastically populated IMF.

\subsection{Tracking high-mass stars}

One of the key new features of this prescription is the tracking of individual high-mass star data. For every star particle, we now have a list of the masses of every star above 8~M$_\ssun$ residing within it. This opens up new science avenues that were not available before.

For example, cosmological simulations may be used to estimate the size and evolution of HII regions, as in \citet{Anderson2017}, based on the ionizing photon output from star particles. Rather than assume ionizing photon production from an SSP, this new prescription allows us to use the specific stars to determine the photon rate, which will add variability to our predictions.

Predictions for stellar remnants, such as pulsar counts, can now be directly inferred from star particles in our simulations. Further, now that we have entered the era of gravitational wave astronomy \citep{Abbott2016}, simulations can be used to predict the merger rates of binary compact objects. This has so far been accomplished by pairing simulation outputs with population synthesis models \citep[e.g.][]{Schneider2017, Mapelli2017, O'Shaughnessy2017, Chakrabarti2017}. Having precise high-mass star information can allow us to refine such predictions by using the actual high-mass star counts from the simulations. Additionally, since the Milky Way has accreted many small galaxies over its lifetime, we can still see the imprint of these single rare events. For example, in the Milky Way stellar halo, accreted UFD-like galaxies may have contributed over half of r-process enhanced metal-poor stars \citep{Brauer2019}. Such predictions, however, are highly sensitive to the actual numbers of massive binary stars in small populations.

\section{Summary}\label{sec:summary}

Motivated by the inability of low-mass halos to self-regulate, and the resulting divergence of different prescriptions in ultra-faint dwarfs, we investigated the treatment of stellar feedback in simulations. In this work, we presented a new treatment of the IMF in cosmological simulations. Informed by observations, we stochastically sample stars from the IMF within each star particle. As a compromise with computational reality, once we have stochastically populated a star particle, we only track individual stars above 8~M$_\ssun$, so that feedback dependent on high-mass stars is calculated for discrete stars, and feedback dependent on low-mass stars is calulated as before, for continuous populations.

To investigate the effects of our new stochastic IMF, we used cosmological zoom-in simulations run to $z=6$ to compare the stellar masses of the resulting ${\sim}100$ galaxies. We found that while galaxies residing in dark matter halos greater than ${\sim}10^{8.5}$~M$_\ssun$ remain unchanged by the new sampling method, galaxies in smaller halos typically have lower stellar masses, by up to an order of magnitude. When comparing the supernova rate via a burstiness parameter, we found that a stochastic IMF leads to significantly burstier feedback because of the greater temporal clustering of supernovae.

To see how the burstier supernova feedback impacts the gas in a galaxy, we used a simulated isolated dwarf galaxy. We ran many versions of both the continuous and stochastically sampled IMF treatment to study systematic differences in star formation, while bracketing scatter between runs. We found that during the first billion years, the galaxy with the stochastic IMF formed as few as half as many stars. The direct cause of this was that as few as half as many gas particles were available to form stars. Feedback with a stochastic IMF is more effective at heating surrounding gas and preventing gas from cooling and condensing into stars.

While another method (``quantized feedback'') is typically used to ensure supernovae are discrete, we found that this method has several inconsistencies, since the actual distribution of high-mass stars is unknown until all supernovae have exploded. We found that the star formation results of quantized feedback were intermediate between a continuous and a stochastic IMF. Quantized feedback leads to ${\sim}10$ and ${\sim}30$ per cent more star formation and available cold gas, respectively. If high-energy radiation contributed energy or momentum in our simulations, it is possible the results would be even more dissimilar.

To test the new IMF prescription, this work focused on galaxies at high redshift. Since the galaxies in the mass range where this prescription is most impactful stop forming stars shortly after reionization, this was sufficient to draw conclusions for faint galaxies. Future work will investigate the affects of a stochastic IMF in simulations run to the present day.

This IMF prescription is ideal for high-resolution simulations; as star particle masses decrease, we can lower the cutoff mass to discretely track lower mass stars, and incorporate discrete treatments for Type Ia explosions and mass loss due to stellar winds. Further, our unique ability to track the evolution of individual stars in cosmological simulations will allow us to make more specific predictions for any observations dependent on the number and distribution of high-mass stars. Future work, for example, can constrain the rates of compact object binary mergers detectable by gravitational wave experiments.

As we explore smaller stellar populations in simulations, we are now afforded the opportunity to investigate astrophysics on smaller scales. With the future predictive power of our simulations in mind, we have implemented a novel stochastic IMF in our cosmological simulations. In future studies of faint galaxies, including their stellar populations and radiative contributions to the epoch of reionization, it will be necessary to use such a stochastically populated IMF to accurately model these phenomena. Otherwise, observational predictions will systematically overestimate the star formation in ultra-faint dwarf galaxies.

\section*{Acknowledgments}

The authors would like to thank the anonymous referee for useful comments that improved the manuscript. We would also like to thank Ferah Munshi and Michael Tremmel for useful discussions relating to this work. EA and AMB  acknowledge support from NSF grant AST-1813871. Resources supporting this work were provided by the NASA High-End Computing (HEC) Program through the NASA Advanced Supercomputing (NAS) Division at Ames Research Center. EA acknowledges support from the National Science Foundation (NSF) Blue Waters Graduate Fellowship. This research is part of the Blue Waters sustained-petascale computing project, which is supported by the National Science Foundation (awards OCI-0725070 and ACI-1238993) and the state of Illinois. Blue Waters is a joint effort of the University of Illinois at Urbana-Champaign and its National Center for Supercomputing Applications.




\bibliographystyle{mnras}
\bibliography{biblist} 




\appendix

\section{Cutoff mass}\label{app:cutoff}

\begin{figure}
    \centering
    \includegraphics[width=0.45\textwidth]{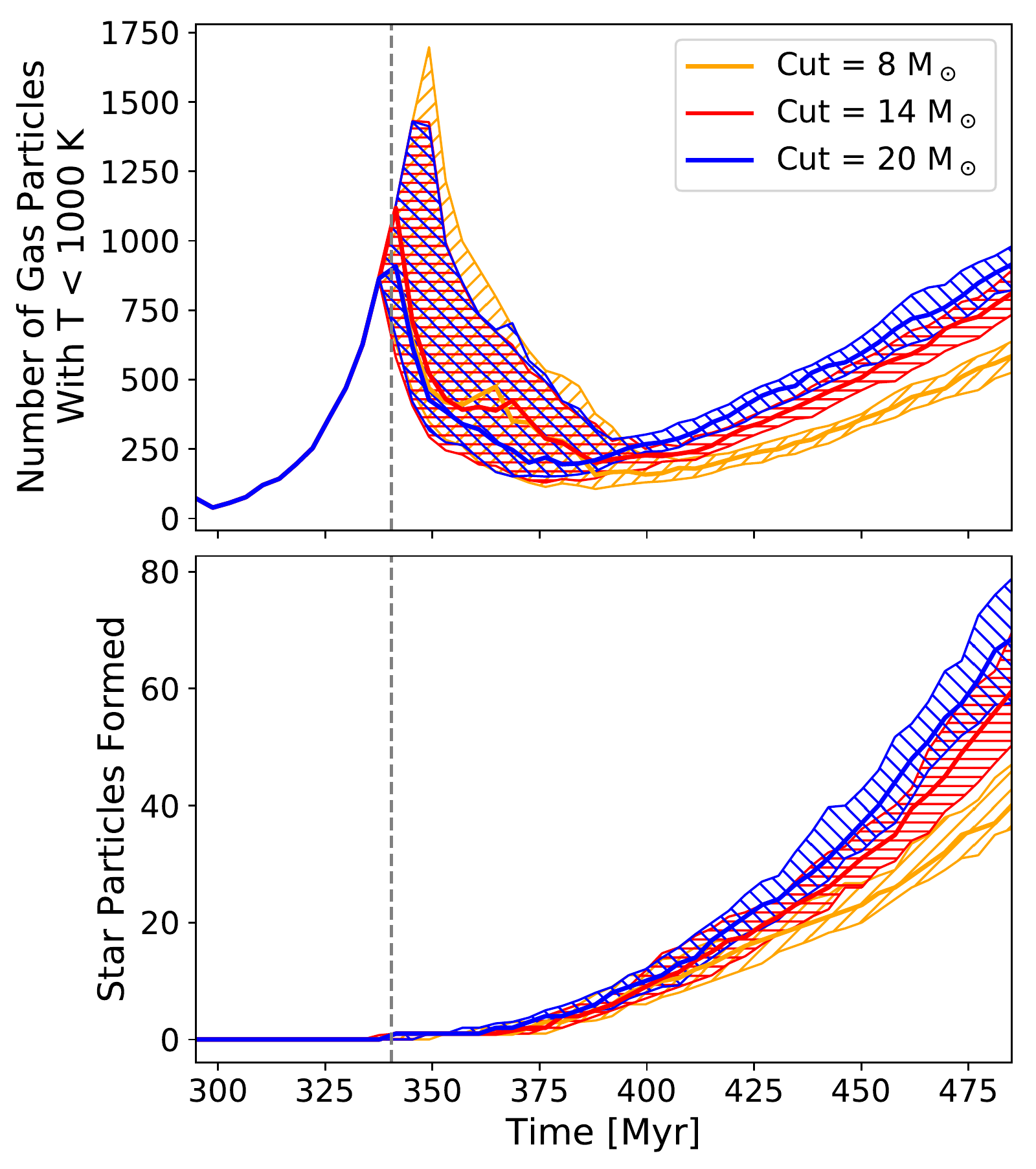}
    \caption{The same as Figs.~\ref{fig:numtemp} and ~\ref{fig:numtempquant}, but instead comparing three versions of the stochastic IMF with different cutoff masses. The higher the cutoff, and therefore the more supernovae are approximated as continuous, the more cold gas is available to form stars and the more stars form. It is therefore important to capture all core collapse supernovae discretely in any IMF sampling method.}
    \label{fig:stochcut}
\end{figure}

It is important to capture all Type II supernovae when employing a discretization scheme. That is, shifting the cutoff mass above the minimum core-collapse supernova mass lessens the impact of the discretization. Fig.~\ref{fig:stochcut} shows this, where we have plotted the first ${\sim}100$~Myr of star formation of 50 runs each of the isolated dwarf galaxy. As we shift the cutoff higher, thereby approximating more of the Type II supernovae as continuous rather than discrete, we approach the fully continuous case of greater star formation and more cold gas. Thus, in order to capture the full impact of the energy deposition, \textit{all} supernovae must be described within the framework of a stochastic IMF.

\section{Numerical considerations}\label{app:numerical}

Here we briefly discuss the computational considerations of applying the IMF sampling method presented in this work. Compared to quantized sampling, this method clearly requires more memory allocated per star particle (specifically, an array containing the list of masses of discrete massive stars tracked). For this reason, this method is best applied only in very high resolution cosmological simulations. For most codes which already associate many pieces of information with their star particles (e.g. temperature, chemical composition, formation time, mass, etc.), a reasonable constraint is to no more than double the memory required per star particle, or equivalently (for typical cosmological codes) to limit the maximum number of individual stars expected during sampling to be of order 10. For this reason, this method is intended for use in simulations with star particle masses ${\lesssim}1000$~M$_\ssun$.

In terms of computation time, sampling in this way is more intensive than quantized feedback or a method akin to \citet{Gatto2017} in which the IMF is sampled only in the high-mass regime. However, as discussed in Section~\ref{sec:stoch}, sampling instead over the entire regime imposes no restriction on the mass or number of stars in the high-mass regime, and so allows for greater variation in both. Fortunately, though the sampling takes time, it is negligible compared to the computation time involved in other aspects of the simulation. Timing comparisons of simulations suggest differences between the two methods are at no more than the percent level.



\bsp	
\label{lastpage}
\end{document}